# Distinct forms of resonant optimality within insect indirect flight motors

Arion Pons[1,2] and Tsevi Beatus[1,2*]

[1] The Silberman Institute of Life Sciences, Hebrew University of Jerusalem, Giv'at Ram, Jerusalem, Israel
[2] The Benin School of Computer Science and Engineering, Hebrew University of Jerusalem, Giv'at Ram, Jerusalem, Israel
* tsevi.beatus@mail.huji.ac.il

**Abstract:** Insect flight motors are extraordinary natural structures that operate efficiently at high frequencies. Structural resonance is thought to play a role in ensuring efficient motor operation, but the details of this role are elusive. While the efficiency benefits associated with resonance may be significant, a range of counterintuitive behaviours are observed. In particular, the relationship between insect wingbeat frequencies and thoracic natural frequencies are uncertain, with insects showing wingbeat frequency modulation over both short and long timescales. Here, we offer new explanations for this modulation. We show how, in linear and nonlinear models of an indirect flight motor, resonance is not a unitary state at a single frequency; but a complex cluster of distinct and mutually-exclusive states, each representing a different form of resonant optimality. Additionally, by characterising the relationship between resonance and the state of negative work absorption within the motor, we demonstrate how near-perfect negative work absorption can be maintained over significant wingbeat frequency ranges. Our analysis leads to a new conceptual model of flight motor operation: one in which insects are indifferent to their precise wingbeat frequency, and robust to changes in thoracic and environmental properties – illustrating the extraordinary robustness of these natural motors.

## 1. Introduction

Insect flight motors are extraordinary: insects show the highest recorded relative locomotion velocities for any animal, reaching up to one thousand body-lengths per second [1]; and can exhibit wingbeat frequencies in the hundreds of Hz [2]. This level of flight motor performance has not been surpassed by any of the wide range of biomimetic analogues under development [3, 4]. Insect flight motors take a range of different forms: one form which has seen detailed study is the indirect flight motor, as found in orders *Diptera* (flies) and *Lepidoptera* (butterflies, moths), as well as in a range of species across other orders [2, 5, 6]. In the indirect flight motor, the contraction of the primary flight muscles – the dorsoventral (DVM) and dorsolongitudinal (DLM) muscles – is transmitted to wingbeat motion via the elastic thorax. This process is thought to involve wing-thorax resonance [7–10] (the absorption of wing inertial loads via thoracic elasticity), but details are elusive. While there is evidence for the significance of resonant effects [11, 12], including consistent wingbeat frequency changes in response to wing mass changes [13], several counterintuitive behaviours are observed. For instance, species that are thought to utilise resonant effects are observed to undergo additional notable, but seemingly random, wingbeat frequency variation – for instance, variation of up to 15% in fruit flies [14]. A range of these species are also observed to alter wingbeat frequency in a controlled manner: in honeybees, fruit flies, and hawkmoths, as a mechanism of aerodynamic force control [9, 15–17]; and in mosquitos, for acoustic courtship interactions [18, 19]. More complex relationship are also observed: both positive and negative [20, 21] correlations between wingbeat frequency and ambient temperature in species of bee and fly, with population-averaged frequency alterations of up to 30% [21]. These behaviours are counterintuitive in the sense that they defy classical conceptions of resonance as a unitary state, existing at a specific frequency. Either insects often deviate from resonance, or more complex effects are at work.

In this vein, several explanations for these behaviours have been proposed. The operation of sets of pleurosternal and tergopleural muscles have been proposed as a mechanism for time-varying control of the thoracic resonant frequency [9, 22–26], potentially allowing wingbeat frequencies to vary while maintaining a state of continued resonance. Alternately, behaviour involving intentional frequency control has been motivated in terms of the trade-off between deviating from resonance, and achieving other goals: for instance, performing a manoeuvre [9], or engaging in courtship [18]. One of the impediments to understanding these effects, and candidate explanations, is the lack of any clear information on what resonance actually represents in an insect indirect flight motor – and what costs, exactly, would be associated with deviating from resonance. It is currently unclear exactly what forms of optimality (*e.g.*, in displacement, velocity, energy, *etc.*) flight motor resonance can represent; which frequencies at which these optimal states occur; and how structural properties (*e.g.*, elasticity distribution within the thorax, and aero-structural damping) affect these frequencies.

In this work, we study these aspects of indirect flight motor resonance in more detail, and obtain some surprising results. We illustrate how there may be up to eight distinct and mutually-exclusive resonant frequencies in the flight motor system – both, classical forms of resonance; and the novel form of global resonance [27–30], defining energetic optimality. Using data from a range of insect species, we demonstrate how the flight motor elasticity distribution, and damping, alter these resonant states. These fundamental theoretical results shed light on discrepancies between wingbeat frequencies and measured thoracic resonant frequencies [11]. They also offer explanations for counterintuitive insect behaviours: how insects can apparently afford to be indifferent to moderate variations in wingbeat frequency [9, 14–16]; and how this indifference may not require time-varying control of the thoracic resonant frequency. The existence of these distinct flight motor resonant frequencies has significant implications both for the interpretation of insect flight, and the design of flapping-wing micro-air-vehicles.

## 2. Transfer functions of hybrid PEA-SEA systems
### 2.1. Dynamics of PEA, SEA and hybrid systems
Insect indirect flight motors are complex dynamical structures [10, 31]. Following Lynch et al. [10], we will develop a new hybrid model for these motors, but consider first the two key ingredients. Consider two distinct linear single-degree-of-freedom (1DOF) forced oscillators: one undergoing parallel-elastic actuation (PEA) and one undergoing series-elastic actuation (SEA). The distinction between these two systems (Fig. 1) is the spatial distribution of elasticity: under PEA, elasticity lies in parallel with the actuator; under SEA, it lies in series (*i.e.*, classical base-excitation) [32]. In both systems, the actuator effect can be expressed in two equivalent forms: in actuator displacement, $u(t)$, or in actuator load, $F(t)$. The equations of motion (EOM) for these systems can be represented:

$$\text{PEA:} \quad m\ddot{x} + d\dot{x} + k_p x = F(t), \qquad u(t) = x(t), \tag{1}$$
$$\text{SEA:} \quad m\ddot{x} + d\dot{x} + k_s x = k_s u(t), \qquad F(t) = k_s\big(u(t) - x(t)\big).$$

Or, in their canonical forms:

$$\begin{aligned}&\text{PEA:}\quad \ddot{x}+2\zeta_p\omega_{0,p}\dot{x}+\omega_{0,p}^2 x=f(t),\quad u(t)=x(t),\\&\text{SEA:}\quad \ddot{x}+2\zeta_s\omega_{0,s}\dot{x}+\omega_{0,s}^2 x=\omega_{0,s}^2 u(t),\quad f(t)=\omega_{0,s}^2\bigl(u(t)-x(t)\bigr),\end{aligned} \quad (2)$$

$$\omega_{0,p}=\sqrt{k_p/m},\qquad \omega_{0,s}=\sqrt{k_s/m},\qquad f(t)=F(t)/m,$$
$$\zeta_p=d/(2\sqrt{mk_p}),\quad \zeta_s=d/(2\sqrt{mk_s}),$$

with natural frequencies $\omega_{0,p}$, $\omega_{0,s}$, and damping ratios $\zeta_p$, $\zeta_s$. Several properties of these systems can be observed. The PEA system behaves as a classical second-order oscillator under prescribed load, $F(t)$; and the SEA system behaves as such under prescribed displacement, $u(t)$. However, consider the reverse: the PEA system under prescribed $u(t)$ behaves as a kinematic link ($u(t)=x(t)$); and the SEA system under prescribed $F(t)$ is governed by an inelastic second-order system ($m\ddot{x}+d\dot{x}=F(t)$, or $\ddot{x}+2\zeta_s\omega_{0,s}\dot{x}=f(t)$).

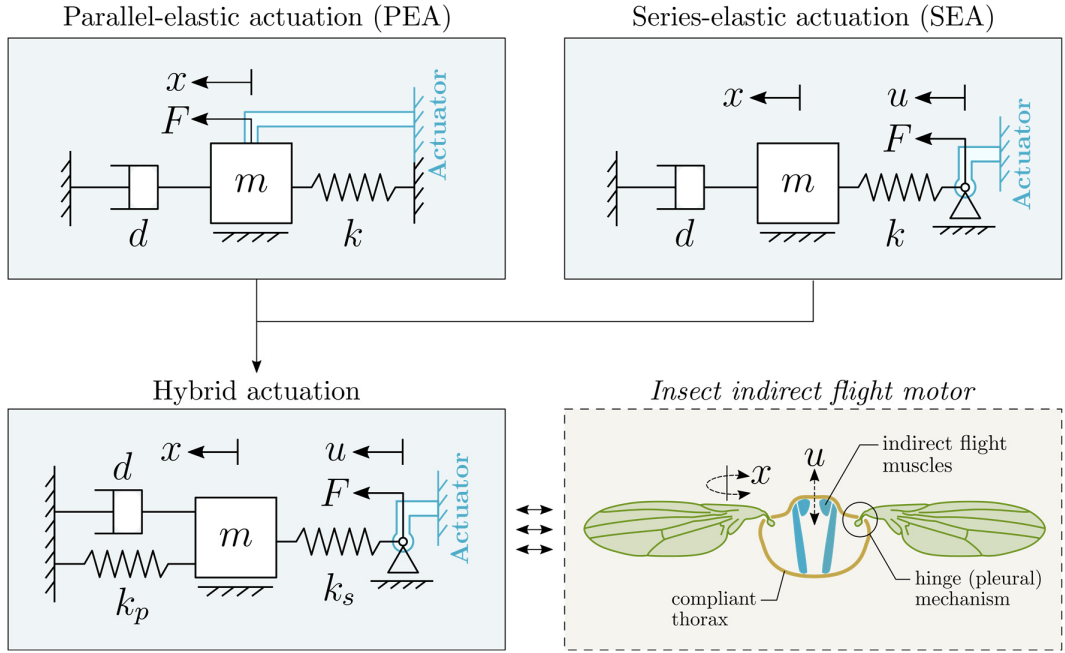

**Fig. 1: Hybrid single-degree-of-freedom system model of the insect indirect flight motor.** Examples of pure parallel-elastic actuation (PEA) and series-elastic actuation (SEA) systems are illustrated, as well as the resulting hybrid model that arises when both a series ($k_s$) and parallel ($k_p$) elasticity is present. This resulting hybrid model can be used as a simple general model of the elasticity distribution within an insect indirect flight motor.

These two systems are simplified models of an insect indirect flight motor (Fig. 1) [10]. They are simplified in that (**i**) our model has linear inertia, elasticity, and damping, rather than any nonlinear form of either [6, 10, 11, 33]; and (**ii**) we have assumed the distribution of elasticity within the motor: pure SEA, or pure PEA. Which distribution of elasticity accurately represents an indirect flight motor? Indirect flight motors are commonly considered to undergo PEA [34–36]. The behaviour of these motors points to dominant parallel-elastic behaviour: to a large degree, insect wings and muscles largely behave as if kinematically connected, as per PEA ($u(t) \approx x(t)$, Eq. 1). Under quasistatic conditions, insect flight muscle deformation is directly coupled to wing motion [7, 11, 37], and induces an elastic response [11]. During flight, the phase difference between muscular deformation and wingbeat motion is often small [5, 31, 38, 39]. However, recent studies [10] have contemplated an additional series-elastic element in indirect flight motors, to account for inexactness in this kinematic connection: the possibility of a phase difference between $u(t)$ and $x(t)$. In some species, *e.g.*, fruit flies [39], a phase difference between muscular contraction and wingbeat motion is observed, pointing to the possible existence of this series-elastic effect. There is physical rationale for considering this series-elastic effect: the transmission of motion from musculature to wing occurs through elastic thoracic structures [7, 10, 31], which have their own local modal frequencies; in principle, introducing series-elastic effects.

Consider, therefore, a new hybrid PEA-SEA model of an insect flight motor (Fig. 1). We have a choice in how to represent the system EOM: with actuation under $u(t)$ or under $F(t)$, that is:

$$m\ddot{x} + d\dot{x} + k_p x = F(t), \text{ or,}$$
$$m\ddot{x} + d\dot{x} + (k_s + k_p)x = k_s u(t), \quad (3)$$
$$\text{with } F = k_s\big(u(t) - x(t)\big)$$

While these two representations are equivalent, the distinction is more than one of notation. Fundamentally, the system's dynamic response to prescribed displacement, $u(t)$, differs from its response to prescribed load, $F(t)$ – just as this response differs in pure PEA and SEA systems. This effect is further elucidated by the system's canonical form:

$$\ddot{x} + 2\zeta_p \omega_{0,p}\dot{x} + \omega_{0,p}^2 x = f(t),$$
$$\ddot{x} + 2\zeta_p \omega_{0,p}\dot{x} + (\omega_{0,s}^2 + \omega_{0,p}^2)x = \omega_{0,s}^2 u(t), \quad (4)$$
$$\text{with } f(t) = \omega_{0,s}^2\big(u(t) - x(t)\big),$$

and with parameters identical to those of the pure PEA and SEA systems (Eq. 2). We note that $\zeta_p \omega_{0,p} = \zeta_s \omega_{0,s} = d/(2m)$ (Eq. 2). As can be seen from Eq. 4, depending on the actuator behaviour (prescribed $f(t)$, or prescribed $u(t)$), the system response is governed by either the purely-parallel natural frequency, $\omega_{0,p}$, or a hybrid natural frequency, dependent on both $\omega_{0,p}$ and $\omega_{0,s}$. In physical terms: in the limit of low damping, $\omega_{0,p}$ represents the system's natural frequency for oscillations in which $u$ is free; and $\omega_{0,s}$, the natural frequency for oscillations in which $u$ is held fixed. To study the system behaviour under stronger damping, we now characterise these natural frequencies in more detail.

### 2.2. Distinct transfer functions and resonant frequencies
The linearity of this hybrid system permits a frequency-domain analysis. We represent the system's time-varying parameters as complex exponentials, or phasors:

$$x(t) = \mathfrak{Re}\{\hat{x}\exp(i\Omega t)\},$$
$$u(t) = \mathfrak{Re}\{\hat{u}\exp(i\Omega t)\}, \quad (5)$$
$$f(t) = \mathfrak{Re}\{\hat{f}\exp(i\Omega t)\},$$

where $\hat{x}, \hat{u}, \hat{f} \in \mathbb{C}$ are complex-valued amplitudes, $\Omega$ is the input/output frequency, and $\mathfrak{Re}\{\cdot\}$ is the real part. These complex-valued amplitudes are equivalent to a real-valued amplitude and phase angle:

$$x(t) = |\hat{x}|\,\mathfrak{Re}\{\exp(i\Omega t + \psi_x)\}, \quad \psi_x = \tan^{-1}\left(\frac{\mathfrak{Im}\{\hat{x}\}}{\mathfrak{Re}\{\hat{x}\}}\right), \tag{6}$$

where $|\hat{x}| \in \mathbb{R}$ is the real amplitude, and $\psi_x \in \mathbb{R}$ is the phase angle. The complex components in Eq. 5 allow us to define two key complex-valued transfer functions; between input and output displacements ($\hat{x}/\hat{u}$) and input load and output displacement ($\hat{x}/\hat{f}$):

$$\frac{\hat{x}}{\hat{u}} = \frac{\omega_{0,s}^2}{\omega_{0,s}^2 + \omega_{0,p}^2 - \Omega^2 + 2i\zeta_p\omega_{0,p}\Omega}, \quad \frac{\hat{x}}{\hat{f}} = \frac{1}{\omega_{0,p}^2 - \Omega^2 + 2i\zeta_p\omega_{0,p}\Omega}, \tag{7}$$

These complex-valued transfer functions can be separated into components of magnitude:

$$\frac{|\hat{x}|}{|\hat{u}|} = \frac{\omega_{0,s}^2}{\sqrt{\left(\omega_{0,s}^2 + \omega_{0,p}^2 - \Omega^2\right)^2 + 4\zeta_p^2\omega_{0,p}^2\Omega^2}},$$

$$\frac{|\hat{x}|}{|\hat{f}|} = \frac{1}{\sqrt{\left(\omega_{0,p}^2 - \Omega^2\right)^2 + 4\zeta_p^2\omega_{0,p}^2\Omega^2}}, \tag{8}$$

and phase:

$$\psi_x - \psi_u = -\tan^{-1}\left(\frac{2\zeta_p\omega_{0,p}\Omega}{\omega_{0,s}^2 + \omega_{0,p}^2 - \Omega^2}\right),$$

$$\psi_x - \psi_f = -\tan^{-1}\left(\frac{2\zeta_p\omega_{0,p}\Omega}{\omega_{0,p}^2 - \Omega^2}\right). \tag{9}$$

We can then probe two key properties of the system: at what frequencies are the response magnitudes, $|\hat{x}|/|\hat{u}|$ and $|\hat{x}|/|\hat{f}|$, maximised? Via the first derivative test, we compute these two distinct frequencies as:

$$\arg\max_\Omega\left(\frac{|\hat{x}|}{|\hat{u}|}\right) = \sqrt{\omega_{0,p}^2 + \omega_{0,s}^2 - 2\omega_{0,p}^2\zeta_p^2} = \omega_{0,p}\sqrt{1 + \alpha^2 - 2\zeta_p^2},$$

$$\arg\max_\Omega\left(\frac{|\hat{x}|}{|\hat{f}|}\right) = \omega_{0,p}\sqrt{1 - 2\zeta_p^2}, \tag{10}$$

where $\alpha = \omega_{0,s}/\omega_{0,p}$,

and $\arg\max_\Omega(H(\Omega)) \equiv \{\Omega : H(\Omega) = \max_\Omega(H(\Omega))\}$.

In Eq. 10 we have introduced the parameter $\alpha$, which conveniently represents systems that are neither pure SEA ($\omega_{0,p} = 0$ and $\omega_{0,s} > 0$, so $\alpha \to \infty$) nor pure PEA ($\omega_{s,0} \to \infty$ and $\omega_{0,p} > 0$, so $\alpha \to \infty$: both pure cases alias onto the same $\alpha$).

These two frequencies are not the only system resonant frequencies. Following the definition of resonance as a state of peak response in some state variable [40], we can define transfer functions in velocity (phasor amplitude $i\hat{x}\Omega$) and acceleration (phasor amplitude $-\hat{x}\Omega^2$):

$$\frac{i\hat{x}\Omega}{\hat{u}} = \frac{i\omega_{0,s}^2 \Omega}{\omega_{0,s}^2 + \omega_{0,p}^2 - \Omega^2 + 2i\zeta_p \omega_{0,p}\Omega}, \quad \frac{i\hat{x}\Omega}{\hat{f}} = \frac{i\Omega}{\omega_{0,p}^2 - \Omega^2 + 2i\zeta_p \omega_{0,p}\Omega},$$

$$\frac{-\hat{x}\Omega^2}{\hat{u}} = \frac{-\omega_{0,s}^2 \Omega^2}{\omega_{0,s}^2 + \omega_{0,p}^2 - \Omega^2 + 2i\zeta_p \omega_{0,p}\Omega}, \quad \frac{-\hat{x}\Omega^2}{\hat{f}} = \frac{-\Omega^2}{\omega_{0,p}^2 - \Omega^2 + 2i\zeta_p \omega_{0,p}\Omega}, \tag{11}$$

The peaks of these transfer functions are the velocity and acceleration resonant frequencies, which are distinct from the displacement resonant frequencies [40]. These four resonant frequencies are:

$$\arg\max_{\Omega}\left(\frac{|i\hat{x}\Omega|}{|\hat{u}|}\right) = \omega_{0,p}\sqrt{1+\alpha^2},$$

$$\arg\max_{\Omega}\left(\frac{|\hat{x}\Omega^2|}{|\hat{u}|}\right) = \omega_{0,p}\frac{1+\alpha^2}{\sqrt{1+\alpha^2 - 2\zeta_p^2}},$$

$$\arg\max_{\Omega}\left(\frac{|i\hat{x}\Omega|}{|\hat{f}|}\right) = \omega_{0,p},$$

$$\arg\max_{\Omega}\left(\frac{|\hat{x}\Omega^2|}{|\hat{f}|}\right) = \frac{\omega_{0,p}}{\sqrt{1-2\zeta_p^2}}. \tag{12}$$

There is now a total of six resonant frequencies in the system. In general, these six frequencies are distinct, as each depend on damping (via $\zeta_p$) and the strength of series-elastic effects (via $\alpha$) in different ways. Fig. 2 illustrates these dependencies. Several points can be noted.

(**i**) The existence of these distinct frequencies raises a point not previously been recognised in the study of insect flight. Even in our simple single-degree-of-freedom linear model, there are multiple distinct resonant frequencies: multiple mutually-exclusive states of optimal actuation. This has implications for the study of indirect flight motor resonance: for instance, existing estimates of honeybee thoracic resonant frequency [11] are based on actuation point acceleration ($\ddot{u}$) with respect to load ($f$). The distinction between this frequency and other resonant frequencies may contribute to observed wingbeat frequency inconsistencies [11].

(**ii**) In terms of the input function variable: broadly, the force-based transfer functions ($\hat{x}/\hat{f}$, *etc.*) would appear to be more relevant to insect flight, as the flight motor metabolic cost is related to its force output [41]. The displacement-based transfer functions ($\hat{x}/\hat{u}$, *etc.*) are only likely to be relevant in cases of extreme performance: where metabolic cost is irrelevant, and the insect's only objective is to extract maximum response from the limited range of muscular motion. It is possible that this situation could be realised in certain extreme evasive or corrective manoeuvres [42–44].

(**iii**) In terms of the output function variable: broadly, the displacement-output ($\hat{x}/\hat{f}$, $\hat{x}/\hat{u}$) and velocity-output ($i\hat{x}\Omega/\hat{f}$, $i\hat{x}\Omega/\hat{u}$) transfer functions would appear to be the more relevant to insect flight, as these output variables are directly related to the insect's aerodynamic performance (via the wingbeat amplitude and peak velocity).

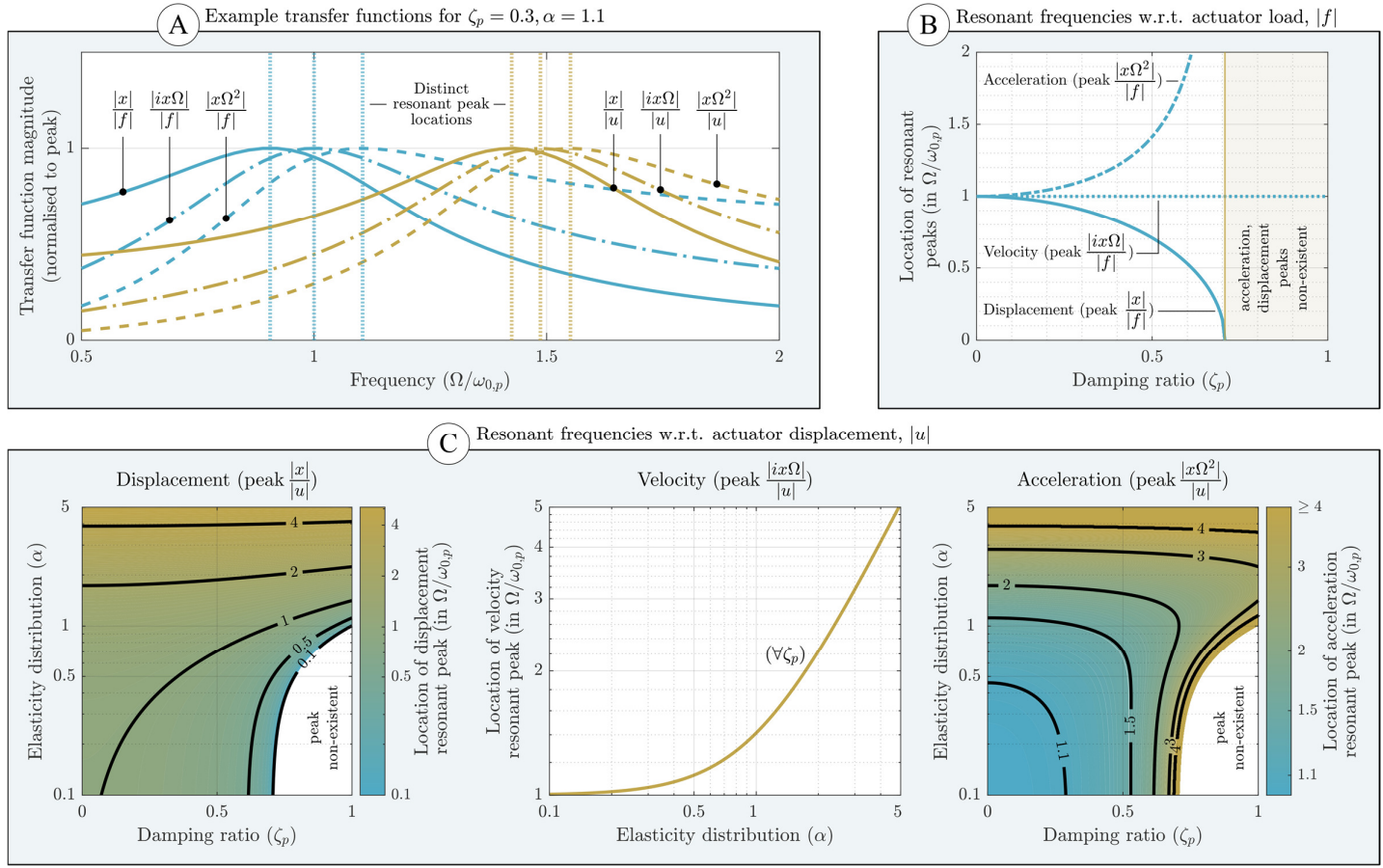

**Fig 2. Six distinct resonant classical resonant frequencies of the insect flight motor model**. (B) An illustration of the six different transfer function magnitudes (Eqs. 7, 8, 11) and the distinct nature of the resonant peaks. Parameters are roughly illustrative of a fruit fly flight motor (*cf.* Sections 4.1-4.2). (**B**) Resonant frequencies with respect to actuator load, $|f|$, *i.e.*, PEA-type resonant frequencies (Eq. 10, 12). (**C**) Resonant frequencies with respect to actuator displacement, $|u|$, *i.e.*, SEA-type, or base excitation-type, resonant frequencies (Eq. 10, 12).

## 3. Power consumption and the global-resonant frequency

### 3.1. Power consumption

The six resonant frequencies defined thus far are not a complete characterisation of the optimal operating points of our simplified hybrid model of an indirect flight motor. We have not yet characterised the flight motor power consumption, and the frequencies at which the flight motor is most efficient. The role of thoracic elasticity, and structural resonance, in reducing motor muscular power consumption can be seen through the lens of negative work absorption: the potential for thoracic elasticity to absorb negative work – work done by the wing on the flight motor – and then release this stored energy at other points during the wingbeat cycle [14, 45–47]. To compute the frequency at which this effect is optimal, we require a more sophisticated analysis. Defining a canonical mechanical power requirement, $p = f\dot{u}$, we find that, as a product of two phasors, $p$ is not representable as a pure phasor [48]. Instead, $p$ is composed of a phasor term and a constant:

$$p(t) = \hat{p}_0 + |\hat{p}|\,\mathfrak{Re}\{\exp(2i\Omega t + \psi_p)\}, \tag{13}$$

With some manipulation, we obtain explicit forms of these phasor and constant terms. First, we represent $f(t)$ and $\dot{u}(t)$ in their analytic forms, $\mathbb{f}(t)$ and $\dot{\mathbb{u}}(t)$:

$$\begin{aligned}f(t) &= \mathfrak{Re}\{\,\mathbb{f}(t)\,\}, & \mathbb{f}(t) &= \hat{f}\exp(i\Omega t),\\ \dot{u}(t) &= \mathfrak{Re}\{\,\dot{\mathbb{u}}(t)\,\}, & \dot{\mathbb{u}}(t) &= \hat{\dot{u}}\exp(i\Omega t) = i\Omega\hat{u}\exp(i\Omega t),\end{aligned} \tag{14}$$

where $\hat{f}$ and $\hat{u}$ are complex-valued amplitudes, encoding both magnitude and phase (Eqs. 5-6). As a reference, we take $\hat{f}$ as a phasor with zero phase ($\hat{f} \in \mathbb{R}$), without loss of generality. We compute $\hat{u}$ in terms of $\hat{f}$ via a chain of the transfer functions defined in Eq. 8:

$$\hat{u} = \left(\frac{\hat{x}}{\hat{u}}\right)^{-1}\left(\frac{\hat{x}}{\hat{f}}\right)\hat{f}. \tag{15}$$

Then we compute $p(t)$ using an appropriate product of its analytic forms [49]:

$$p(t) = \mathfrak{Re}\{\dot{\mathbb{u}}\mathbb{f}\} = \underbrace{\frac{1}{4}(\dot{\mathbb{u}}^\dagger\mathbb{f} + \dot{\mathbb{u}}\mathbb{f}^\dagger)}_{\hat{p}_0} + \underbrace{\frac{1}{4}(\dot{\mathbb{u}}\mathbb{f} + (\dot{\mathbb{u}}\mathbb{f})^\dagger)}_{|\hat{p}|\,\mathfrak{Re}\{\exp(2i\Omega t + \psi_p)\}}, \tag{16}$$

where $(\cdot)^\dagger$ denotes the complex conjugate. As noted in Eq. 17, terms in this product can be identified with the constant ($\hat{p}_0$) and oscillatory ($|\hat{p}|$) components of the time-domain signal. The constant terms are given by the associated complex product:

$$\frac{\hat{p}_0}{\hat{f}^2} = \frac{\zeta_p \omega_{0,p} \Omega^2}{\left(\omega_{0,p}^2 - \Omega^2\right)^2 + 4\zeta_p^2\omega_{0,p}^2\Omega^2}, \tag{17}$$

The time-domain terms are given by the coefficients of the complex exponentials in the complex product:

$$\begin{aligned}\frac{1}{4}(\dot{\mathbb{u}}\mathbb{f} + (\dot{\mathbb{u}}\mathbb{f})^\dagger) &= \hat{p}^+\exp(2i\Omega t) + \hat{p}^-\exp(-2i\Omega t),\\ |\hat{p}| &= \sqrt{(\hat{p}^+)^2 + (\hat{p}^-)^2},\end{aligned} \tag{18}$$

with magnitude, $|\hat{p}|$, given by:

$$\frac{|\hat{p}|}{\hat{f}^2} = \frac{\Omega}{2\alpha^2\omega_{0,p}^2} \sqrt{\frac{\left((1+\alpha^2)\omega_{0,p}^2 - \Omega^2\right)^2 + 4\zeta_p^2\omega_{0,p}^2\Omega^2}{\left(\omega_{0,p}^2 - \Omega^2\right)^2 + 4\zeta_p^2\omega_{0,p}^2\Omega^2}}. \qquad (19)$$

The parameters $\hat{p}_0$ and $|\hat{p}|$ define the power consumption of our hybrid model of an indirect flight motor (Eq. 13). With them, we can characterise the state of negative work absorption in the motor.

### 3.2. Global resonance and optimal efficiency

The insight that thoracic elasticity can absorb flight motor negative work, and thereby decrease motor muscular power consumption [14, 45–47] is connected to a more general phenomenon of linear and nonlinear dynamics: the generalised energy-based conception of resonance sometimes referred to as *global resonance* [27–30]. In a general system, the core condition of global resonance is the absence of negative work in the system overall power requirement:

$$p(t) \geq 0, \forall t. \qquad (20)$$

If this condition is satisfied – *e.g.*, at a certain frequency, or state of elasticity – then the system is optimally efficient in a particular sense that we now study. Consider the overall power consumption of our indirect flight motor model, measured by the integrals:

$$\bar{p}_{\text{abs}} = \frac{1}{T}\int_0^T |p(t)|\, dt, \qquad \bar{p}_{\text{pos}} = \frac{1}{T}\int_0^T p(t)[p(t) \geq 0]_{\mathbb{I}}\, dt, \qquad (21)$$

where $T = 2\pi/\Omega$ and $[\,\cdot\,]_{\mathbb{I}}$ is the Iverson bracket[1] [50]. The metric $\bar{p}_{\text{abs}}$ is the absolute power consumption [51, 52]. It describes the behaviour of an actuator without energy regeneration, in which both positive ($p > 0$) and negative work requirements ($p < 0$, *e.g.*, for braking wingbeat motion) must be provided at full energetic cost. The metric $\bar{p}_{\text{pos}}$ is positive-only power [45, 46]. It describes the behaviour of an actuator with imperfect energy regeneration, in which a specialised dissipative element is available to absorb, but not store, negative work.

Exactly which of these metrics best describes the flight motor musculature is unclear. However, it is clear at least that this musculature does not show perfect energy regeneration, *i.e.*, the ability to innately convert negative work back to metabolic energy. In fact, muscles often consume energy when doing negative work [53]. In any case, under global resonance, *i.e.*, the absence of negative work (Eq. 20), $\bar{p}_{\text{abs}}$ and $\bar{p}_{\text{pos}}$ take identical values. More formally, we can define the mechanical power transfer efficiencies:

$$\eta_{\text{abs}} = \frac{\int_0^T p(t)\, dt}{\int_0^T |p(t)|\, dt}, \qquad \eta_{\text{pos}} = \frac{\int_0^T p(t)\, dt}{\int_0^T p(t)[p(t) \geq 0]_{\mathbb{I}}\, dt}, \qquad (22)$$

representing the ratio of actual net power throughput to actuator power consumption ($\bar{p}_{\text{abs}}$, $\bar{p}_{\text{pos}}$). We can see that the maximum possible efficiency is $\eta_{\text{abs}} = \eta_{\text{pos}} = 1$, and that this condition is reached only in the absence of negative work, *i.e.*, at global resonance (Eq. 20).

---

[1] $[\lambda]_{\mathbb{I}} = 1$ for true statement $\lambda$, and $[\lambda]_{\mathbb{I}} = 0$ for false statement $\lambda$.

We seek, therefore, the frequencies at which our hybrid model of an indirect flight motor is global-resonant. Denoting this frequency $\omega_g$, we can translate Eq. 20 to a condition on the power offset $\hat{p}_0$ and amplitude $|\hat{p}|$:

$$\omega_g = \{\Omega : \hat{p}_0 \geq |\hat{p}|\} \qquad (23)$$

Analysing Eq. 24 in detail, via Eqs. 17 and 19, we can confirm that there exists no $\Omega$ such that $\hat{p}_0 > |\hat{p}|$ strictly: the equation $\hat{p}_0 = |\hat{p}| + \epsilon$ has no real solutions in $\Omega$ for $\epsilon > 0$. Therefore, the condition for no negative work is uniquely:

$$\omega_g = \{\Omega : \hat{p}_0 = |\hat{p}|\}, \qquad (24)$$

which we can solve explicitly by equating Eqs. 17 and 19. Note first that, for pure PEA ($\omega_{0,s} \to \infty$) and pure SEA ($\omega_{0,p} = 0$), we have simply:

$$\begin{array}{ll} \text{PEA:} & \text{SEA:} \\ \omega_{0,s} \to \infty, & \omega_{0,p} = 0, \\ \omega_g = \omega_{0,p}, & \omega_g = \omega_{0,s}\sqrt{1 - 4\zeta_s^2}, \end{array} \qquad (25)$$

We pause briefly on this result. In a pure PEA system, the global resonant frequency is precisely the undamped natural frequency of the system: a seemingly fundamental link, but it does not generalise to other actuator configurations. In the pure SEA system, the global resonant frequency involves the unique factor $4\zeta_s^2$, distinct from all other resonant frequencies. And in the full hybrid case, the solution to Eq. 24 is the root of a quartic polynomial:

$$\omega_g^\pm = \omega_{0,p}\sqrt{1 + \delta \pm \sqrt{\delta^2 - 4\zeta_p^2}}, \quad \delta = \frac{1}{2}\alpha^2 - 2\zeta_p^2. \qquad (26)$$

These $\omega_g^\pm$ are nonclassical resonant frequencies which, to our knowledge, have not previously been studied. Their behaviour is illustrated in Fig. 3. We note:

(**i**) The hybrid system has up to *two* global resonant frequencies, $\omega_g^-$ and $\omega_g^+$, associated with PEA- and SEA-type behaviour, respectively. These frequencies are distinct from the classical resonant frequencies (Eqs. 10, 12), and, when real-valued, are always $\geq \omega_{0,p}$.

(**ii**) There exist some $\alpha$ and $\zeta_p$ for which $\omega_g^\pm$ are complex-valued. In this situation, true global resonance does not exist. The efficiencies ($\eta_{abs}, \eta_{pos}$) may still have maxima, but not at $\eta_{abs} = \eta_{pos} = 1$: we term these maxima pseudo-global-resonant states. Further analysis is required to determine the exact frequency of these pseudo-global-resonant states, but we propose an approximation: $\mathfrak{Re}\{\omega_g^\pm\}$. Consider the classical complex-valued transfer functions (Eqs. 7, 11): the natural frequency, $\omega_{0,p}$, is the real part of the transfer function poles, and thus approximates their resonant peak locations. In the same way, the real part of the transfer function efficiency poles approximates their pseudo-global-resonant peak locations.

(**iii**) Irrespective of whether true global resonance exists or not, a state of near-global resonance, where $\eta_{abs} \approx 1$, is available over a comparatively wide range of frequencies. As Fig. 3B illustrates, the efficiency profiles are notably flat-topped, compared to equivalent classical transfer functions (Fig. 2). For parameters roughly indicative of a fruit fly flight motor (Sections 4.1-4.2), a state of <1% inefficiency ($\eta_{abs} \geq 99\%$), is satisfied over a frequency band of width $0.31\omega_{0,p}$. In classical transfer functions, a comparable state (magnitudes $\geq 99\%$ of peak value) is available only over windows of width $\approx 0.09\omega_{0,p}$. Physically, this is significant: an indirect flight motor only needs to operate in the broad vicinity of the global resonant state to attain almost all of its energetic benefit. This provides some explanation for insects' apparent indifference to precise wingbeat frequency [9, 14–16], as we discuss in Section 5.

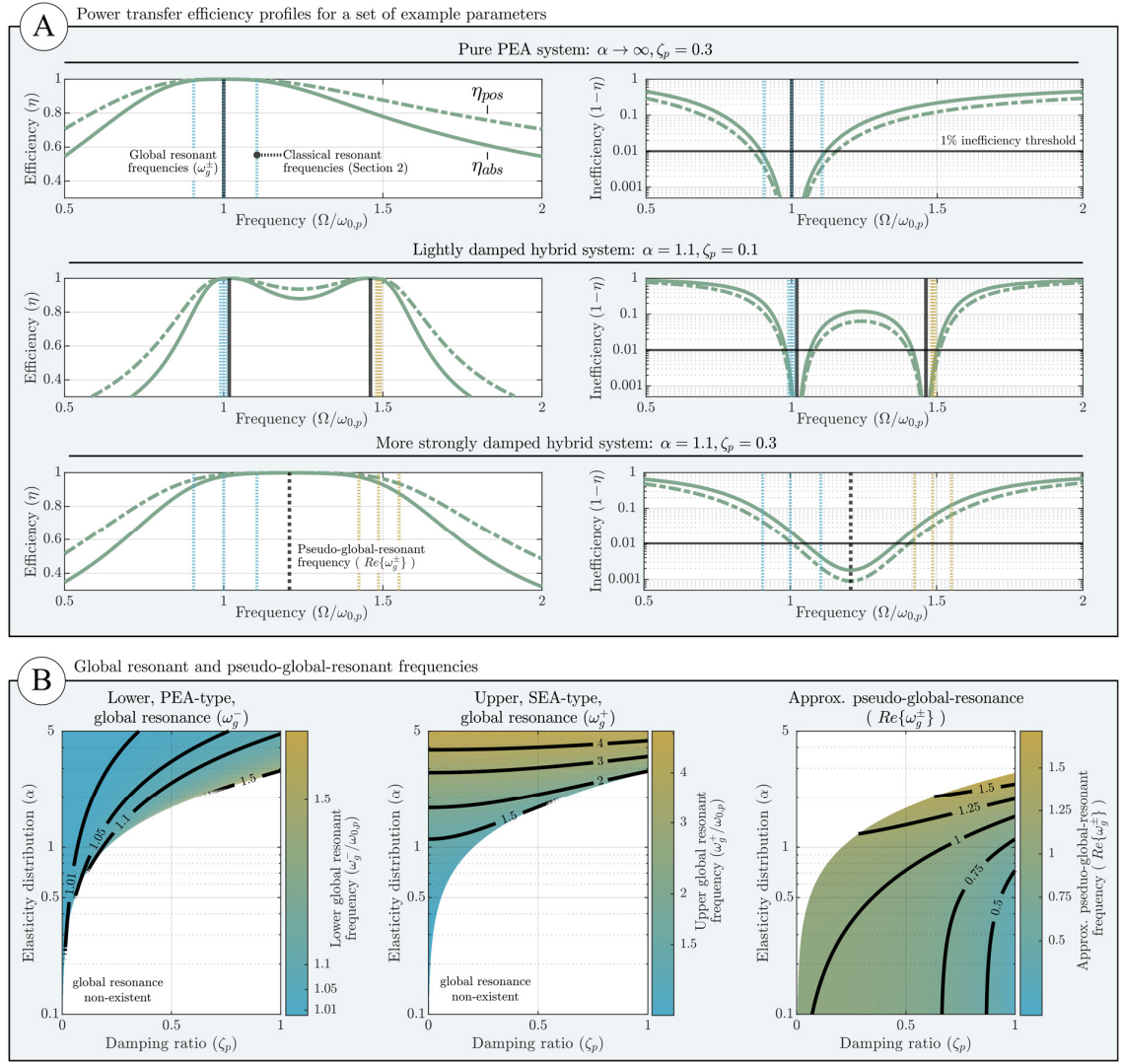

**Fig 3. Global resonance of the hybrid system.** (**A**) The power transfer efficiencies, $\eta_{\text{abs}}$ and $\eta_{\text{pos}}$, over frequency, $\Omega$, for a set of example parameter values. These examples illustrate the rich behaviour of the hybrid system under global resonance. In a pure PEA system, $\omega_g^\pm = \omega_0$, however, in the hybrid system (e.g., $\alpha = 1.1$) the two global resonant frequencies may be distinct, or they may not exist at all. The latter is the case for a more strongly damped hybrid system – though behaviour that is very close to global resonant ($\eta_{\text{abs}} > 99\%$) is available over a wide range of frequencies. (**B**) The global resonant frequencies of the hybrid system, as a function of $\alpha$ and $\zeta_p$. Alongside the true global resonant frequencies, $\omega_g^\pm$, we describe an approximate pseudo-global-resonant frequency ($\Re\mathfrak{e}\{\omega_g^\pm\}$): the approximate location of the efficiency maxima, when these maxima are not $\eta_{\text{abs}} = \eta_{\text{pos}} = 1$, i.e., true global resonance is not available.

## 4. Effects of aerodynamic nonlinearity
### 4.1. Dynamics of systems with quadratic nonlinearity

We have thus far considered a linear model of an insect indirect flight motor. In reality, insect flight motors show many different forms of nonlinearity, including nonlinear aerodynamic damping [10, 33], and nonlinear structural elasticity [6, 11]. The linear model allowed us to illustrate the distinction between different forms of resonance in an analytical framework – and indeed, even this linear system stretched the limits of analytical computation with the root of a quartic in Eq. 26. The distinction between resonant states generalises to the nonlinear case: nonlinear resonant states are also distinct, and in broadly the same way. To demonstrate this, and to illustrate the possible effects of nonlinearity, we perform a brief numerical study. Consider a pure PEA model of an indirect flight motor with quadratic damping:

$$\begin{aligned}\text{dimensional:} \quad & m\ddot{x} + c|\dot{x}|\dot{x} + k_p x = F(t), \\ \text{canonical:} \quad & \ddot{x} + \beta|\dot{x}|\dot{x} + \omega_{0,p}^2 x = f(t), \quad \beta = c/m,\end{aligned} \quad (27)$$

as might better represent the effect of nonlinear aerodynamic drag [10]. Assume that the output response of the system is simple-harmonic: $x(t) = \hat{x}\cos(\Omega t)$, as might approximate an insect wingbeat [33]. Normalising this response in amplitude and time, as $x(t) = \hat{x}q(\Omega t) = \hat{x}q(\tau)$, yields the complete dimensional reduction:

$$\begin{aligned}\tau = \Omega t, \quad \Lambda = \Omega/\omega_{0,p}, \quad q(\tau) = x(\tau)/\hat{x}, \quad v(\tau) = f(\tau)/\hat{x}\omega_{0,p}^2 \\ \implies \Lambda^2(q'' + \beta\hat{x}|q'|q') + q = v(\tau) \equiv V[q(\tau)],\end{aligned} \quad (28)$$

where $(\cdot)'$ denotes $d/d\tau$. An important property is observed in Eq. 28: the system dynamics, under this prescribed output response, are not governed by $\beta$ and $\hat{x}$ independently, but only by the product $\beta\hat{x}$. This property significantly simplifies the analysis. Expressing the relationship in Eq. 28 as a functional, $v(\tau) = V[q(\tau)]$, we can define nonlinear transfer ratios relating the peak output response (in non-normalised $x, \dot{x}, \ddot{x}$) to the peak load requirement:

$$\begin{aligned}\frac{\max_t|x|}{\max_t|f|} &= \frac{\hat{x}}{\hat{x}\omega_{0,p}^2}\frac{\max_\tau|q|}{\max_\tau|v|} = \frac{1}{\omega_{0,p}^2\max_\tau|V[\cos(\tau)]|}, \\ \frac{\max_t|\dot{x}|}{\max_t|f|} &= \frac{\hat{x}\Omega}{\hat{x}\omega_{0,p}^2}\frac{\max_\tau|q'|}{\max_\tau|v|} = \frac{\Lambda}{\omega_{0,p}\max_\tau|V[\cos(\tau)]|}, \\ \frac{\max_t|\ddot{x}|}{\max_t|f|} &= \frac{\hat{x}\Omega^2}{\hat{x}\omega_{0,p}^2}\frac{\max_\tau|q''|}{\max_\tau|v|} = \frac{\Lambda^2}{\max_\tau|V[\cos(\tau)]|}.\end{aligned} \quad (29)$$

Note that these transfer ratios capture only a segment of this nonlinear oscillator's behaviour: its behaviour when outputting simple-harmonic waves. The maximisation in Eq. 29 can be performed numerically. Three dimensionally-reduced resonant frequencies for this nonlinear system can then be computed, as a function of the damping product parameter ($\beta x$):

$$\begin{aligned}\Lambda_{r,x}(\beta\hat{x}) &= \arg\max_\Lambda\left(\frac{1}{\max_\tau|V[\cos(\tau)]|}\right) \\ \Lambda_{r,\dot{x}}(\beta\hat{x}) &= \arg\max_\Lambda\left(\frac{\Lambda}{\max_\tau|V[\cos(\tau)]|}\right),\end{aligned} \quad (30)$$

$$\Lambda_{r,\ddot{x}}(\beta\hat{x}) = \arg\max_{\Lambda}\left(\frac{\Lambda^2}{\max_{\tau}|V[\cos(\tau)]|}\right).$$

Continuing in this harmonic-output framework, we seek to compute the frequency at which this nonlinear system is global-resonant. The system power requirement can be expressed in an analogous functional form:

$$\begin{aligned} p &= f\dot{x} = \hat{x}^2\omega_{0,p}^2\Omega v q', \text{ and } \mathcal{P}[q(\tau)] = V[q(\tau)]q', \\ &\implies p = \hat{x}^2\omega_{0,p}^2\Omega \cdot \mathcal{P}[q(\tau)]. \end{aligned} \quad (31)$$

Using $\mathcal{P}[q(\tau)]$, we can compute power transfer ratios defined in Eq. 22 (via numerical integration), and the frequencies at which these transfer ratios take their maxima (via numerical optimisation). Formally, for $\eta_{\text{abs}}$:

$$\begin{aligned} \eta_{\text{abs}}(\beta\hat{x}, \Lambda) &= \frac{\int_0^T p(t)\,dt}{\int_0^T |p(t)|\,dt} = \frac{\int_0^{2\pi} \mathcal{P}[\cos(\tau)]\,d\tau}{\int_0^{2\pi} |\mathcal{P}[\cos(\tau)]|\,d\tau} \\ \Lambda_g(\beta\hat{x}) &= \arg\max_{\Lambda}\bigl(\eta_{\text{abs}}(\beta\hat{x}, \Lambda)\bigr). \end{aligned} \quad (32)$$

In general, these resonant states may be true global resonant (peak $\eta_{\text{abs}} = 1$) or pseudo-global-resonant (peak $\eta_{\text{abs}} < 1$).

### 4.2. Distinction between resonant frequencies under aerodynamic nonlinearity

Fig. 4 illustrates these resonant frequencies over a representative range of the product parameter $\beta\hat{x}$, with transfer functions and ratios illustrated for example parameter values. Overall, the behaviour of this nonlinear system is strikingly similar to the linear PEA model. Several points may be noted:

(**i**) Regarding the distinctness of frequencies. Over the studied range, the acceleration resonant frequency ($\Lambda_{r,\ddot{x}}$) is always equal to the undamped natural frequency ($= 1$). The global resonant frequency, $\Lambda_g$, is also always equal to the undamped natural frequency ($= 1$) and it is always true global resonant (peak $\eta_{\text{abs}} = 1$). Numerically, these equivalences are confirmed to within machine precision. The displacement and velocity resonant frequencies, $\Lambda_{r,\dot{x}}$ and $\Lambda_{r,\ddot{x}}$ differ from the undamped natural frequency ($= 1$) and from each other. This distinctness parallels the behaviour of linear PEA systems (Fig. 2), except acceleration resonance, rather than velocity resonance, coincides with the undamped natural frequency. The resonant frequencies are also ordered in the same way ($\Lambda_{r,\ddot{x}} \geq \Lambda_{r,\dot{x}} \geq \Lambda_{r,x}$).

(**ii**) Regarding the variation in frequencies. The velocity and displacement resonant frequencies show a strong decreasing trend with $\beta\hat{x}$. The effect of decreasing resonant frequency with increasing damping (analogous to the linear PEA system) becomes more pronounced with increasing amplitude, as per the analogy of a variable linear damping coefficient that depends on velocity ($d = c\hat{x}\Omega$). This would suggest the existence of a damping-driven coupling effect between wingbeat amplitude and resonant frequency in insects: an effect that appears in *in vitro* experimentation (Section 5.3).

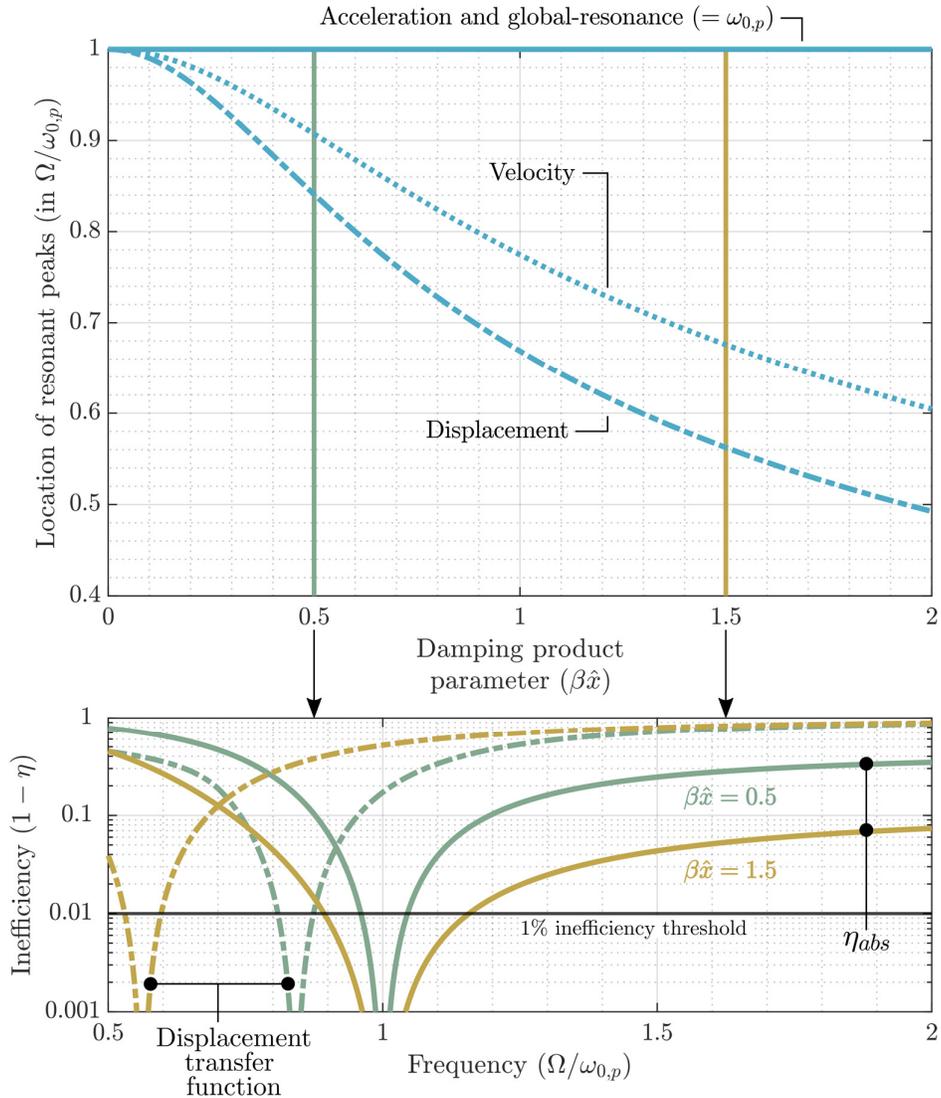

**Fig 4. Resonant properties of a PEA system with quadratic damping.** Resonant frequencies (as per Eqs. 30, 32) for a PEA system with quadratic damping, as a function of the damping product parameter ($\beta\hat{x}$). These frequencies pertain to the case in which the oscillator generates a simple-harmonic output, and define frequencies of minimum peak load requirement (and, for global resonance, the state of no negative work). Transfer functions and transfer ratios are illustrated for two example product parameter values ($\beta\hat{x} = 0.5, 1.5$).

## 5. Practical application to insect flight

### 5.1. Identifying damping parameters

To utilise the results of Sections 2-4 to characterise flight motor resonance in specific insect species, model parameters must be identified: in particular, the overall motor damping ($\zeta_p$, or $\beta$), and the strength of the series-elastic effect ($\alpha$). A range of disparate data sources can be drawn upon for parameter estimation. We begin with a characterisation of motor damping. As studied in Section 4, insect flight motor damping is nonlinear. To best represent the overall energetic effect of nonlinearity in our linear model, we use estimates of the quality factor ($Q$- factor) for insect flight. Using the $Q$-factor, we can translate nonlinear system information (energy loss per cycle) to linear model properties (approximate damping ratio, $\zeta_p$). Two approaches for empirical $Q$-factor estimation have been presented in the literature: estimates based on flight motor energetics [54] or empirical correlations based on the Weis-Fogh number [10]. Reported $Q$-factor estimates based on these two approaches show significant inconsistency: estimates, *e.g.*, for bumblebees range from $Q = 1.6$ [10] to $Q = 19$ [54]. We note that this inconsistency is due to a factor four calculation error in current energetic estimates [54]. As a corrected calculation, consider an energetic definition of the $Q$-factor of a linear oscillator [54]:

$$Q = 2\pi \frac{K}{\Delta E}, \tag{33}$$

where $\Delta E$ is the oscillator energy loss per cycle, and $K$ is the peak potential energy (= peak kinetic energy). To compute $Q$ using energetic parameters, two values are used [54]: $\overline{P}_{aero}$, the mean wing aerodynamic power dissipation, and $\overline{P}_{acc}$, the mean inertial power requirement over a single wingbeat quarter-cycle. To relate these semi-empirical terms to $K$ and $\Delta E$, note that: (**i**) the relationship between $\Delta E$ and $\overline{P}_{aero}$ is simply the definition of the mean [54]:

$$\Delta E = T\overline{P}_{aero} = \frac{2\pi}{\Omega}\overline{P}_{aero}, \tag{34}$$

and, (**ii**) the relationship between $K$ and $\overline{P}_{acc}$ is given by the definition of $\overline{P}_{acc}$ [54]:

$$\overline{P}_{acc} = \frac{K}{\frac{1}{4}T} = \frac{2}{\pi}\Omega K, \tag{35}$$

that is, $\overline{P}_{acc}$ integrated over a quarter-cycle yields the peak system kinetic energy. The consistent estimate of $Q$, based on $\overline{P}_{acc}$ and $\overline{P}_{aero}$, is thus:

$$Q = \frac{\pi}{2}\frac{\overline{P}_{acc}}{\overline{P}_{aero}}, \tag{36}$$

This yields a revised estimate of $Q = 1.6$ for fruit flies [14]; $Q = 2.5$ for hawkmoths, and $Q = 4.7$ for bumblebees [35]. These values are consistent with those computed via Weis-Fogh number correlations [10]: $Q = 1.2$ for fruit flies; $Q = 1.7$ for hawkmoths; and $Q = 1.6$ for bumblebees. To translate these $Q$-factors to estimates of $\zeta_p$, we assume $\Omega \approx \omega_{0,p}$, implying that [55] $Q \approx 1/2\zeta_p$. This yields estimates of $0.31 \leq \zeta_p \leq 0.42$ for fruit flies; $0.20 \leq \zeta_p \leq 0.29$ for hawkmoths; and $0.11 \leq \zeta_p \leq 0.31$ for bumblebees. Via Weis-Fogh number correlation [10, 36], we can extend this analysis to different species, noting, *e.g.*, an estimate of $N = 6.3$, $Q = 2.2$, $\zeta_p = 0.22$ for honeybees. Results from computational fluid dynamics [45] may be able to refine these estimates via more precise estimates of the energy loss per cycle ($\Delta E$).

### 5.2. Identifying series-elastic parameters

The second parameter we require an estimate of is the strength of the series-elastic effect ($\alpha$). We can identify this parameter using measurements of the phase difference across the motor, *e.g.*, between muscular contraction and wingbeat motion. In our hybrid model, this phase difference is uniquely defined by $\alpha$, $\zeta_p$, and $\Omega/\omega_{0,p}$ (Eq. 9). With a phase difference estimate $\Delta\psi = \psi_x - \psi_u$, we compute the parameter $\varepsilon = -1/\tan(\psi_x - \psi_u)$. Based on the assumed operating point of the flight motor ($\Omega/\omega_{0,p}$), we can identify $\alpha$. For instance, at different resonant states (Eqs. 10, 12, 26):

Displacement resonance:
$$\Omega = \omega_{0,p}\sqrt{1 - 2\zeta_p^2},$$
$$\alpha^2 = 2\zeta_p^2\left(1 + \varepsilon\sqrt{1 - 2\zeta_p^2}\right),$$

Velocity resonance:
$$\Omega = \omega_{0,p},$$
$$\alpha^2 = 2\zeta_p\varepsilon,$$

Acceleration resonance:
$$\Omega = \frac{\omega_{0,p}}{\sqrt{1 - 2\zeta_p^2}},$$
$$\alpha^2 = \frac{2\zeta_p\left(-\zeta_p + \varepsilon\sqrt{1 - 2\zeta_p^2}\right)}{1 - 2\zeta_p^2},$$

Global resonance: (37)
(Eq. 26)
$$\alpha^2 = \frac{2\zeta_p(1 + \varepsilon^2)\left(\zeta_p \pm \sqrt{\varepsilon^2 + \zeta_p^2}\right)}{\varepsilon^2}$$

These values provide bounds on $\alpha$, assuming that the flight motor is operating at a frequency bounded by these four frequencies. In practical terms, the velocity resonance / undamped natural frequency ($\omega_{0,p}$) estimate serves as a useful simple indicator of whether series-elastic effects are significant.

To estimate motor phase differences, a range of sources, including microCT [31], X-ray diffraction [38, 39], and laser profilometery [5] studies, are available. Fig. 5 shows two representative analysis processes:

(**i**) For bumblebees (*Bombus* spp.) [38]: X-ray diffraction data provides estimates of flight muscle sarcomere length variation under tethered flight. This variation is well-described by sinusoidal fits (Fig. 5A). Referencing these fits against approximate wingbeat cycle tracking [38], we estimate a mean phase lag of $\Delta\psi = -3.2°$ ($\varepsilon = 17.9$). This implies that series-elastic effects are insignificant (at $Q = 3.2$, $\zeta_p = 0.16$, and so $\alpha = 2.4$). With both low damping and weak series-elastic effects, resonant frequencies are clustered close together: within a window of 6% of $\omega_{0,p}$. Nevertheless, the frequency variation available under <1% inefficiency is broad: a window of 12% of $\omega_{0,p}$. Even in insects with relatively undamped flight motors, near-perfect elastic energy absorption is likely to be available over a significant frequency window.

(**ii**) For fruit flies (*Drosophila mettleri*) [39]: X-ray diffraction data provides estimates of flight muscle sarcomere length variation under tethered flight (Fig. 5B). The time resolution of this data is relatively coarse, but maxima can be identified, as per [39], and phase lags estimated: a mean of $\Delta\psi = -22.0°$ ($\varepsilon = 2.47$). Using damping estimates from other *Drosophila* species ($Q = 1.4$, $\zeta_p = 0.36$ [10, 14]), we estimate $\alpha = 1.3$. Series-elastic effects are significant, and resonant frequencies are widely dispersed: classical force-based frequencies lie within a window of 30% of $\omega_{0,p}$, and a pseudo-global-resonant frequency exists at approximately $1.28\omega_{0,p}$. Based on the computed efficiency profiles, the frequency variation available under <1% inefficiency is very broad: a window of 54% of $\omega_{0,p}$. Even if this system had no series-elastic effects ($\zeta_p = 0.36$, $\alpha \to \infty$), this window would still be 27% of $\omega_{0,p}$. This is illustrative of how, in insects with more strongly damped indirect flight motors, elastic energy absorption is not a practical limit on frequency variation. There are other factors, including the tuning of the passive wing hinge [43], which may limit the available frequency variation: but absorption of negative work is unlikely to be one of them.

For other species, data is less complete. For blowflies [31], MicroCT visualisation shows phase lags of no more than a few degrees between bulk thoracic motion and wingbeat motion, indicating that series-elastic effects are insignificant. Damping estimates are still required. For hawkmoths [5]: significant lateral wingbeat phase differences are noted, but symmetric thorax-wing phase differences are not recorded. The existence of significant lateral phase differences indicates that series-elastic effects may be significant, but further data is needed.

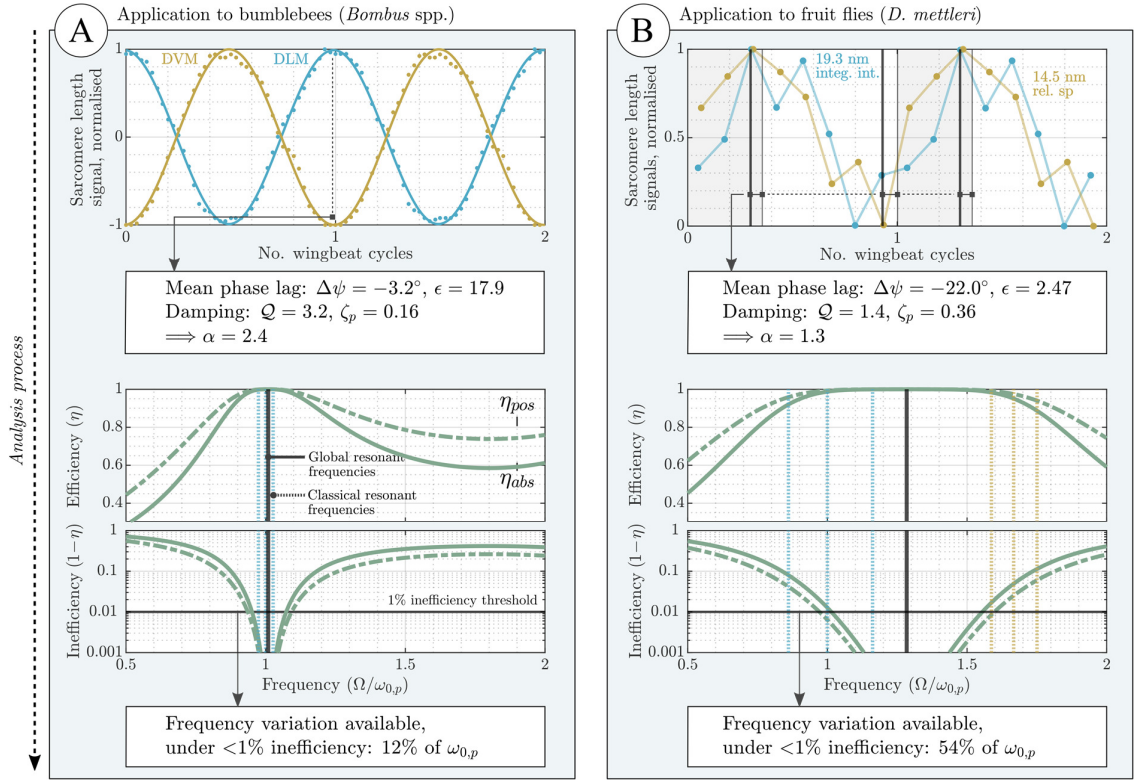

**Fig 5. Application of our analysis process to different insect species.** (**A**) Bumblebees, *Bombus* spp., with X-ray diffraction data extracted from [38]; (**B**) Fruit flies, *Drosophila mettleri*, with representative X-ray diffraction data extracted from [39] (14.5 nm relative spacing, 19.3 nm integrated intensity). In both cases, X-ray diffraction allows estimates of the phase lag ($\Delta\psi$) between muscular and wingbeat oscillation during tethered flight. This information, combined with damping estimates (Section 5.1) allows an estimate of the strength of the series-elastic effect ($\alpha$), and the wingbeat frequency variation available while maintaining <1% mechanical inefficiency (Section 3.2). Even in the relatively lightly-damped bumblebee, this variation is significant: 12% of the natural frequency, $\omega_{0,p}$.

### 5.3. Implications for the experimental measurement of thoracic resonant frequency

The distinction between resonant frequencies outlined in Sections 2-3 has experimental ramifications. Consider recent estimates of honeybee thoracic resonant frequency: an average resonant frequency of 368 Hz [11], relative to a species wingbeat frequency of 220-250 Hz [56]. The resonant frequency measured is the peak of the transfer function between actuation acceleration ($\ddot{u}$) and actuation load ($f$). As per Sections 2-4, several factors could lead to a difference between this frequency and other system resonant frequencies: linear damping ($\zeta_p$); series-elastic effects ($\alpha$), and nonlinear damping ($\beta\hat{x}$). Regarding linear damping: the representative transfer function shown in [11] has a full-width at half-maximum of 34 Hz. This corresponds to a $Q$-factor of 12.2 [10], or $\zeta_p = 0.04$: significantly lower than in-flight damping estimates ($\zeta_p = 0.22$, Section 5.1). At $\zeta_p = 0.04$, damping-based distinctions between resonant frequency are small (Eqs. 10, 12). Assuming therefore that $\omega_{0,p} \approx 368$ Hz, we can estimate in-flight motor resonant frequencies: based on Eq. 10, we estimate the displacement resonant frequency to be 5% below $\omega_{0,p}$, at 350 Hz.

The discrepancy between experimental ($\zeta_p = 0.04$) and in-flight ($\zeta_p = 0.22$) damping levels is qualitatively well-explained by nonlinear aerodynamic damping (Section 4), with the increase in effective linear damping arising from increasing wingbeat amplitude (via $\beta\hat{x}$). Additional information, *e.g.*, simultaneous wingbeat tracking, is needed for quantitative predictions; but the existence of this effect supports the prediction of damping-driven nonlinear coupling between wingbeat amplitude and thoracic resonant frequency. Regarding series-elastic effects: in computing these damping-based estimates, we have assumed these effects are insignificant, in the absence of available data. If series-elastic effects are significant, then interpreting these experimental results is much more complex. In our hybrid linear model, the transfer function between $\ddot{u}$ ($-\Omega^2\hat{u}$) and $f$ is:

$$\frac{-\Omega^2\hat{u}}{\hat{f}} = -\Omega^2 \left(\frac{\hat{x}}{\hat{u}}\right)^{-1} \left(\frac{\hat{x}}{\hat{f}}\right). \tag{38}$$

Even in this simple linear system, the expression for the resonant frequency of this transfer function is complex, involving the root of a general quartic polynomial. In species in which series-elastic effects may be significant, *e.g.*, in fruit flies, a reliance on actuation point input/output transfer functions may not be feasible: simultaneous wingbeat tracking may be required to accurately measure thoracic resonant frequencies.

### 5.4. Implications for wingbeat frequency variation in insects
Finally, we note that the distinctness between resonant frequencies in our flight motor model provides two possible explanations for observed insect wingbeat frequency variation. Firstly, with reference to flight control: it is possible that insects prioritise particular resonant states (*e.g.*, peak energetic efficiency *vs.* peak performance) under particular conditions. This would represent controlled variation in wingbeat frequency to achieve particular distinct optimal states – a feature which could relate to wingbeat frequency modulation as a mechanism of flight control [9, 15, 16]. Gau et al. [9] observe frequency variation windows of up to 32% w.r.t. the mean in hawkmoths; Combes et al. [15], up to 17% in honeybees; Lehmann and Dickinson [16], up to 19% in fruit flies. These frequency changes are broadly consistent with differences between resonant frequencies w.r.t. force ($|\hat{x}|/|\hat{f}|$, *etc.*) and/or a global resonant frequency. It is possible that, in insects with strong series-elastic effects (*e.g.*, fruit flies, Section 5.2), the resonant frequencies w.r.t. displacement ($|\hat{x}|/|\hat{u}|$, *etc.*) could also be involved. A metanalysis of reported wingbeat frequency variation is tabulated in Gau et al. [9]. If these variations represent switches between distinct resonant frequencies, then finely-tuned resonant frequency control via pleurosternal and tergopleural muscular actuation would not be required.

Secondly, with reference to energetic optimality: if, as is often thought [14, 45–47], the purpose of thoracic elasticity is to absorb negative power requirements, then this purpose is fulfilled over frequency ranges, not only at single frequencies. These frequency ranges are sufficient to explain most currently-reported wingbeat frequency variations. Again, for this purpose, finely-tuned resonant frequency control via pleurosternal and tergopleural muscular actuation would not be required. Broadly, our characterisation leads to a new conceptual model of insect wingbeat frequency behaviour. Instead of a model in which insects are bound to a precise wingbeat frequency, determined by structural and environmental parameters and muscular tuning, we propose that insects can be indifferent to the precise match between wingbeat frequency and motor/thoracic resonant frequency, without significant energetic cost. This robustness to frequency mismatch provides additional basis for other forms of flight robustness observed in the literature: robustness to wear [57] damage [8, 58], and environmental and metabolic conditions [17].

## 5. Conclusions

In this work, we have described the distinctness of resonant frequencies within a model of the insect indirect flight motor: a multiplicity of mutually-exclusive resonant states, each of which could represent different optimal operating states. The distinctness of these resonant states alters our understanding of indirect flight motors; sheds new light on counterintuitive insect behaviour; and leads to methods for identifying motor structural parameters from observable data. We introduce several forms of resonance that have not previously been considered in a flight motor context, including global resonance, and illustrate the dependency of these resonant states on flight motor structural parameters, including damping, and the distribution of structural elasticity. Via the analysis of global resonance, we are able to characterise the effect of wingbeat frequency variation on the level of negative work absorption within the motor, and show how near-perfect negative work absorption is maintained over significant frequency windows. These new aspects of flight motor model behaviour potentially explaining wingbeat frequency modulation behaviour observed in many insect species, and contribute to our increasing understanding of the remarkable robustness of insect flight.

**Funding**     This work was supported by the Israel Ministry of Science and Technology, and the Israel Science Foundation (grant number 1851/17). AP was additionally supported by the Jerusalem Brain Community Post-Doctoral Fellowship.